\documentclass[aps,prb,twocolumn,showpacs,superscriptaddress,groupedaddress]{revtex4}  
\usepackage{graphicx}  
\usepackage{dcolumn}   
\usepackage{amsmath}
\usepackage{bm}        
\usepackage{amssymb}   
 \usepackage[usenames]{color}

\usepackage{color}
\newcommand{\ds}{\displaystyle}
\newcommand{\nn}{ \nonumber}
\newcommand{\tr}{{\rm Tr}}

\begin{document}



\title{Quantum thermodynamics for driven dissipative bosonic systems}

\author{Maicol A. Ochoa}
\affiliation{Department of Chemistry, University of Pennsylvania, Philadelphia PA 19104, USA}

\author{Natalya Zymbovskaya}
\affiliation{Department of Physics and Electronics, University of Puerto Rico-Humacao, CUH Station, Humacao, PR 00791,USA}

\author{Abraham Nitzan}
\affiliation{Department of Chemistry, University of Pennsylvania, Philadelphia PA 19104, USA}
\affiliation{School of Chemistry, Tel Aviv University, Tel Aviv 69978, Israel}



\date{\today}

\begin{abstract}
  We investigate two prototypical dissipative bosonic systems under slow driving and arbitrary system-bath coupling strength, recovering their dynamic evolution as well as the heat and work rates, and we verify that thermodynamic laws are respected. Specifically, we look at the damped harmonic oscillator and the damped two-level system. For the former, we study independently the slow time-dependent perturbation in the oscillator frequency and in the coupling strength. For the latter, we concentrate on the slow modulation of the energy gap between the two levels. Importantly, we are able to find the entropy production rates for each case without explicitly defining nonequilibrium extensions for the entropy functional. This analysis also permits the definition of phenomenological friction coefficients in terms of structural properties of the system-bath composite.   
\vspace{2mm}

\end{abstract}

\pacs{
}
\maketitle

\section{Introduction}
 The formulation of thermodynamic concepts applicable to molecular and nanoscale devices has recently motivated intense research, as such systems provide a unique setting to study thermodynamic functions, heat transfer, power work and dissipation at the nanoscale far from the thermodynamic limit. The  characteristics of these systems forbid the direct application of traditional concepts from macroscopic statistical thermodynamics, because fluctuations, thermal and quantum, as well as the system's coupling to its environment can be relevant for their complete description. In the quantum regime, dynamics\cite{Mitchison2015}, broadening of energy levels and interference between different pathways can play important roles and have been studied within the emerging field of quantum thermodynamics\cite{Seifert2012,Horodecki2013,Gelbwaser2015,Ng2015,Vinjanampathy2016,Goold2016,Esposito2015quantum,millen2016perspective}. Models of quantum heat engines that mimic macroscopic setups, for example two-level Otto engines that operate in two/four-stroke and continuous cycles\cite{Uzdin2015,Uzdin2016,Gelbwaser2017single} have been discussed, highlighting the role played by quantum dissipation and friction\cite{zolfagharkhani2005quantum,Wang2007,volokitin2011quantum,intravaia2014quantum,Plastina2014,Alecce2015,Shiraishi2016,klatt2017quantum} and providing frameworks for analysing efficiency and power in quantum heat engines \cite{Curzon1975,Uzdin2014, Rutten2009, Esposito2009,Seifert2011,Izumida2012,Uzdin2014,Whitney2014,Bauer2016}. Recently  a setting for the realization of a four-stroke Otto engine with single trapped ions was theoretically  suggested\cite{Abah2012,Rossnagel2014} and experimentally achieved\cite{Rossnagel2016}. Implications of quantum thermodynamics have also been discussed in the framework of driven open quantum systems as may be encountered in quantum pumps, where the driving appears via suitable time dependence of the system's Hamiltonian. 

Such models are often discussed in the weak system-bath coupling limit, were the thermodynamics functions associated with the system of interest can be clearly identified. In contrast, in the strong coupling limit one encounters difficulties partly stemming from the fact that uncertainty about assigning the system-bath coupling to any part of the overall system, and also because quantum mechanical broadening makes it difficult to exactly characterize the system energy. A simple example is the driven resonant level\cite{Esposito2009Thermoelectric,Kennes2013,Ludovico2014,Proesmans2016,Ludovico2016,Bruch2016,ochoa2016energy}, where a single electronic level is coupled to a Fermi bath (or several such baths) while its energy and/or coupling to the bath are modulated by an external force. In the weak system-bath coupling regime, stochastic thermodynamics\cite{Sekimoto2010,seifert2012stochastic} and (for periodic driving) Floquet theory\cite{kohler2005driven}  have been sucessfully used for describing  transport and thermodynamic implications of such driving in a consistent form\cite{cuetara2015stochastic,Proesmans2016}. Strong system-bath coupling\cite{campisi2009thermodynamics,Dare2016,Esposito2015nature,Carrega2016,ochoa2016energy,jarzynski2017stochastic, perarnau2017fundamental,strasberg2017stochastic}, has proven more challenging (strong coupling in nanothermoelectric devices is discussed in Ref.\  \citenum{Katz2016} ). In another context, the appearance of paradoxical behavior and anomalies in thermodynamic quantities\cite{campisi2010thermodynamic, ingold2012thermodynamic} such as the specific heat\cite{hanggi2008finite,ingold2009specific} has raised questions about the possibility to achieve a consistent thermodynamic description of strongly coupled quantum systems.

The driven resonant level model has been useful for understanding the implications of strong system-bath coupling on the quantum thermodynamics of small systems. In this paper we investigate the quantum thermodynamics of two other prototypical systems operating in the strong coupling regime and under slow driving -- a driven harmonic oscillator and a drvien two level system strongly coupled to their thermal bosonic environments. We aim for a unified description of  dynamic and thermodynamic properties of these systems. The first model Eqs.\ \eqref{eq:Hamil}-\eqref{eq:HamilYm} below, a harmonic oscillator strongly coupled to a bosonic bath, and  driven by modulating in time its characteristic energy, (i.e the oscillator's frequency) or coupling to the bath, may be applied to describe some physical systems such as optomechanical heat engines\cite{zhang2014quantum,dechant2015all}, or molecules adsorbed on insulator surfaces and subjected to mechanical stress. In previous theoretical studies, such models have been used to formulate harmonic quantum Otto engines with time-dependent frequency\cite{Abah2012} as well as other quantum heat engines\cite{lin2003performance, rezek2006irreversible}, and have served to study the interplay between Markovian quantum master equations and Floquet theory\cite{kohler1997floquet} under parametrically periodic driving. Indeed, the forced quantum harmonic oscillator weakly coupled to a thermal bath (the latter modeled as a set of two-level systems) was analyzed using stochastic thermodynamics\cite{horowitz2012quantum}. Recently, experimental studies of the quantum thermodynamics of a two-dimensional quantum harmonic oscillator having angular momentum were reported\cite{de2017experimental} .
 
The second model Eqs.\ \eqref{eq:Hamil_TLS}-\eqref{eq:HintTLS}, a driven dissipative two-level system in the strong coupling regime, is similar to models used in quantum optics and quantum electrodynamics but different from the familiar spin-boson model, (the dynamics for the latter was thoroughly described in, for example, Ref.\ \citenum{Leggett1987}). Previous studies using this model have concentrated on identifying quantum signatures in the thermodynamic behavior of such models in the weak coupling regime\cite{friedenberger2017assessing}. A parametric one-dimensional oscillator in a time-dependent potential has been studied as a dissipative two-level system\cite{zerbe1995brownian}. Studies under strong driving and non-Markovian dynamics\cite{schmidt2015work} stressing the nature of work and heat transfer, quantum jump approximations to the work statistics\cite{hekking2013quantum} and the dynamics and thermodynamics near equilibrium\cite{semin2014nonequilibrium} have been reported. Notably, some experimental aspects associated with measuring work and heat in a dissipative two-level quantum system, where only parts of the system and its environment are accessible to the measurement, were analyzed\cite{viisanen2015incomplete}. 

In contrast to these studies, the present work does not consider sudden {\sl adiabatic steps} that uncouple the original system from the surrounding baths, as such ideal steps may not reproduce important aspect of their practical realization. Indeed, the operation of nanoengines often involves continuous variations such as the migration of chemically bonded molecules on surfaces, plasmon-exciton couplings, optically trapped nanobeads and optical tweezers. Our strategy closely follows  the methodology adopted in Ref.\ \citenum{Bruch2016} in the study of the driven resonant level model, focusing on the dependence of thermodynamic properties of the overall (system + bath) system on system parameters. As a consequence of the bosonic nature of the system under investigation,  we are able to go beyond driving in the oscillator's frequency and consider in addition the time-dependent perturbations on the coupling strength for the damped harmonic oscillator. Moreover, we identify quantum friction terms under finite-rate driving for each case and we achieve a consistent dynamics as well as thermodynamic characterization in each case. 

In Sec. \ref{sec:HO} we study the damped harmonic oscillator exposed to external perturbations that drive the oscillator frequency and the coupling. Next, in Section \ref{sec:TLS}, we describe the thermodynamics of the damped harmonic oscillator when the driving changes the energy gap between levels. These results lead to the subsequent discussion of quantum friction in Sec.  \ref{sec:friction}. We summarize and conclude in Sec. \ref{sec:conclusions}.

\section{ The Driven Damped Harmonic Oscillator}\label{sec:HO}

In this section we study a driven harmonic oscillator coupled to a harmonic bath. The starting point is the standard Hamiltonian (here and below we set $\hbar = 1$)
\begin{align}
\hat H=& \hat H_S+\hat H_B+\hat V \label{eq:Hamil}
\intertext{with}
\hat H_S =& \Omega \hat a^\dagger \hat a,\\
\hat H_B =& \sum_m \omega_m \hat b_m^\dagger \hat b_m,\\
\hat V =& \sum_m u_m \hat X \hat Y_m,\\ 
\hat X =& \hat a+ \hat a^\dagger,\\
\hat Y_m =&\hat b_m^\dagger + \hat b_m,\label{eq:HamilYm}
\end{align}
where, $\hat a$ ($\hat a^\dagger$) is the annihilation ( creation ) operator for the primary boson of frequency $\Omega$, coupled to a bath of bosonic modes of frequencies $ \omega_m $, coupling to the primary boson $ u_m $ and the corresponding annihilation ( creation ) operators $\hat b_m $ ($\hat b_m^\dagger$). This bath is at thermal equilibrium with temperature $T= (k_B \beta)^{-1}$ where $k_B$ is the Boltzmann constant.  

In describing the dynamics of this system, considerable simplication is achieved by resorting to the rotating wave approximation, keeping in Eq.\ \eqref{eq:Hamil} only coupling terms that can conserve energy in low order. In this case the dynamics is fully described by Green functions of the form $\langle \hat a (t) \hat a^\dagger (t') \rangle$.  We define the nonequilibrium Green function in the Keldysh contour 
\begin{align}
G(\tau_1, \tau_2) =& -i \langle \hat a(\tau_1) \hat a^\dagger(\tau_2) \rangle_c \label{eq:GreenF} ,
\end{align}
and notice that the lesser projection $G^<$ at equal times provides the reduced nonequilibrium density matrix for the primary boson, i.e.,
\begin{align}
  \rho(t) =& i G^<(t,t).\label{eq:densitymat}  
\end{align} 

In thermal equilibrium these functions are conveniently described in frequency space.  As in the driven resonance level model (Refs.\ \citenum{Ludovico2014,Bruch2016,ochoa2016energy}), the dynamics of the process under study reflect the fact that upon driving, the system explores different regimes of bath population, the Fermi distribution in Refs.\ \citenum{Bruch2016},\citenum{ochoa2016energy} and the Bose-Einstein distribution here. For simplicity we follow Refs.\ \citenum{Bruch2016},\citenum{ochoa2016energy}  in disregarding other effects, in particular those associated with the bath band structure by invoking the wide band approximation. For static problems this is justified under the assumption that $\Gamma$ is small enough so that its frequency dependence is not explored within the width of the spectral function $A(\omega)$. A necessary condition is that the bath spectral region explored by the system is well above $\omega=0$ and well below any cutoff such as the environmental Debye frequency $\omega_D$ , i.e., $0 \ll \Omega \ll \omega_D$ and $\Gamma \ll \omega_D$. If $\Gamma$ is constant within this regime the retarded projection $G^r$ and the corresponding spectral density (density of modes projected on the primary boson) $A(\omega) = -2 {\rm Im}(G^r(\omega))$ take the form (see Appendix  \ref{ap:GrHO})

\begin{align}
  G^{r}(\omega) =& \frac{1}{\omega - \Omega +i(\Gamma/2)},\label{eq:GrHO_ss}\\
 A(\omega) =& \frac{\Gamma}{(\omega- \Omega)^2+(\Gamma/2)^2},\label{eq:spectralFun}
\end{align}
where (with $g(\omega)$ being the density of modes of the free bath)
\begin{align}
 \Gamma(\omega) =& 2 \pi \sum_k |u_k|^2 \delta(\omega_k - \omega)\\  \label{eq:gamma_def}
  =& \int d\omega_k g(\omega) |u_k|^2 \delta(\omega_k - \omega),
\end{align}
is assumed to be independent of $\omega$. Under these assumptions, the  part of the free energy (the canonical potential) that depends on system parameters ($\Omega$ and $\Gamma$) is given by


\begin{align}
  F(\Omega,\Gamma)=& \frac{1}{\beta}\int_{\omega_o}^\infty \frac{d \omega}{2 \pi} A(\omega) \ln \left(1 - e^{- \beta \omega} \right) \label{eq:GrandPot}.  
\end{align}

In Eq.\ \eqref{eq:GrandPot} $\omega_o >0$ is the cutoff frequency introduced to guarantee that the integral is finite and well-defined. The effect of this lower cutoff on the rates that we evaluate in this section is assessed in Appendix   \ref{ap:Cutoff} and found to be irrelevant for the present analysis as long as $\omega_o$ is smaller than other characteristic energies of the system (i.e. $0< \omega_o \ll \Gamma, \Omega$).  In the following, we omit the limits of integration when writing integrals but we always keep in mind that a lower cutoff $\omega_o$ has been set. The canonical potential $F(\Omega, \Gamma)$ can be used to determine the dependence on system parameters of all other thermodynamic functions relevant to our calculation (see Sec.\ \ref{sec:HO_omega}).

The analysis in subsections \ref{sec:HO_omega} and \ref{sec:HOGamma} below is done under this assumption. It is also possible that $\Gamma$ is small enough to justify the wide band forms \eqref{eq:GrHO_ss} and \eqref{eq:spectralFun} of the Green and spectral functions but is changing as $\Omega(t)$ explores different regimes of the bath spectrum. This case can be treated by assuming that $\Gamma$ is independent of $\omega$ but depends on $\Omega(t)$, see subsection \ref{sec:osc_NWBA}.

In what follows we investigate the effect of driving either on the frequency $\Omega$ or the couplings $u_m$ (and consequently $\Gamma$), limiting our discussion to the case in which local driving is slow compared with the relaxation rate that drives the system into equilibrium. Specifically, we consider that driving in $\Omega$ is slow if the relation $\Omega^{-1} d_t \Omega \ll \Gamma$ holds, and also if $\Gamma^{-1}d_t\Gamma \ll \Gamma_{\rm min}$ is valid when the driving is in the coupling terms $u_k$, with $\Gamma_{\rm min}$ corresponding to the minimum value on $\Gamma$ achieved during modulation.  

\subsection{Driving the oscillator frequency}\label{sec:HO_omega}
The extreme limit where $\Omega$ varies infinitely slowly with time is referred to as the quasistatic limit, where there is a complete timescale separation between the internal system dynamics and the external driving. In this limit all equilibrium relationships remain valid except that $\Omega(t)$ replaces the constant $\Omega$. In the wide band approximation the retarded Green function and correspondng spectral density, Eqs.\ \eqref{eq:GrHO_ss}, \eqref{eq:spectralFun} become

\begin{align}
  G^{r}(t,\omega) =& \frac{1}{\omega - \Omega(t) +i(\Gamma/2)},\label{eq:GrHO_ss_new}\\
 A(t,\omega) =& \frac{\Gamma}{(\omega- \Omega(t))^2+(\Gamma/2)^2},\label{eq:spectralFun_new}
\end{align}
the latter satisfies the following differential property
\begin{align}
  \frac{\partial}{\partial \omega} A(t, \omega) = - \frac{\partial}{\partial \Omega} A(t, \omega).
\end{align}
The canonical potential, Eq.\ \eqref{eq:GrandPot} is given by
\begin{align}
    F(\Omega,\Gamma)=& \frac{1}{\beta}\int_{\omega_o}^\infty \frac{d \omega}{2 \pi} A(t,\omega) \ln \left(1 - e^{- \beta \omega} \right), \label{eq:GrandPot2}
\end{align}
and can be used to find the quasistatic entropy (as before, we focus on the $\Omega$ dependent part of this and all other thermodynamic functions)

 The equilibrium (quasistatic) energy $E^{0}$ for the composite system (primary boson+bath) can be obtained from the canonical potential  $ F $  utilizing the expression  $E^{(0)} = F + T S^{(0)}$, where $S^{(0)}$ represents the absolute entropy of the composite.  Using the canonical potential $ F $ given by Eq.\ \eqref{eq:GrandPot} we compute the corresponding $ \Omega $-dependent contributions to all relevant thermodynamic functions. Thus the entropy $ S^{(0)} $ accepts the form:
\begin{align}
S^{(0)}(t) = & k_B \beta^2 \frac{\partial}{\partial \beta }F
\nn \\  =& 
-k_B \int \frac{d \omega}{2 \pi} A(t, \omega)\Big [n(\omega) \ln n(\omega) 
\notag\\ &
- (1+n(\omega)) \ln (1+ n(\omega))\Big]\label{eq:entropy},
\end{align}
where $n(\omega)$ is the Bose-Einstein distribution $n(\omega) = (e^{\beta \omega}-1)^{-1}$, and the quasistatic energy $E^{(0)}$ and heat capacity $C^{(0)} = (\partial/\partial T) E^{(0)}$
\begin{align}
  E^{(0)}(t) =&F+TS^{(0)}= \int \frac{d\omega}{2 \pi} A(t, \omega) \omega n(\omega)\label{eq:revenergy},\\ 
  C^{(0)}(t) =& k_B \beta^2 \int \frac{d \omega}{2 \pi} \omega^2 A(t,\omega) n(\omega)(1 + n(\omega)). \label{eq:HeatCapacity}
\end{align}

In Eqs.\ \eqref{eq:entropy}, \eqref{eq:revenergy} and  \eqref{eq:HeatCapacity} the superscript $(0)$ indicates that the corresponding quantity does not depend on the rate $\dot \Omega$. It is interesting to notice that these expressions for the equilibrium energy $E^{(0)}$ as well as the heat capacity $C^{(0)}$ suggest that an extended subsystem that includes the primary boson and a fraction of the coupling region will effectively describe the thermodynamics of the full system. To illustrate this point we again focus on that part of the total (system + bath) energy that depends on system parameters and following the methodology in Ref. \citenum{ochoa2016energy}, we extend the definition of the canonical potential in Eq. \eqref{eq:GrandPot} by introducing rescaling parameters which allow for the computation of the independent contributions to the total system-bath energy from the primary boson part $\hat H_S$ , the harmonic bath $\hat H_B$ and the coupling term $\hat V$ (see Appendix \ref{ap:Elambda}). The resulting expressions read
\begin{align}
\langle \hat H_S \rangle =& \Omega \int \frac{d\omega}{2 \pi} A(\omega) n(\omega),\\
\langle \hat V \rangle =& 2 \int \frac{d\omega}{2 \pi} A(\omega) (\omega - \Omega)n(\omega),\\
  \langle \hat H_B \rangle =& -\frac{1}{2} \langle \hat V \rangle.
\end{align}
Consequently $E^{(0)} = \langle \hat H _S \rangle + (1/2)\langle \hat V \rangle$ which suggests that an effective system with Hamiltonian $\hat H _{\rm eff} = \hat H_S + (1/2)\hat V$ defines the extended system. While this result may be appealing, we stress that the occurence of an effective Hamiltonian is not needed in the present discussion of the equilibrium thermodynamics.


 Equivalent expressions can be written in terms of rates. For example, the rate of change of the internal energy $E$ is obtained from Eq.\ \eqref{eq:revenergy} to be
\begin{align}
  \dot E^{(1)} = \dot \Omega \frac{\partial}{\partial \Omega}E^{(0)}, \label{eq:derE1}
\end{align}
where the superscript indicates that this rate is linear in $\dot \Omega$.

The reversible work associated with infinitesimal variations in $\Omega$ must abide to the maximum work principle, therefore $d W = d\Omega \partial_{\Omega} F$. Consequently, the reversible power for quasi-static driving is 
\begin{align}
  \dot W^{(1)} =& \dot \Omega \frac{\partial}{\partial \Omega}F = \dot \Omega \int \frac{d\omega}{2 \pi} A(t, \omega) n(\omega) \label{eq:work_1_HO_omega}.  
\end{align}
 This result indicates that reversible work rate is proportional to the equilibrium population $\langle n \rangle = (2 \pi)^{-1}\int d\omega A(\omega) n(\omega) $ in the primary boson according to  $\dot W = \dot \Omega \langle n \rangle$.

The quasistatic heat generated from an infinitesimal transformation is proportional to the infinitesimal change in the entropy of the system as given by the differential $d Q = d\Omega T \partial_\Omega S$.
Hence
\begin{align}
\dot Q^{(1)} =& \dot \Omega  T \frac{\partial}{\partial \Omega} S^{(0)} = \dot \Omega \int \frac{d \omega}{2 \pi} A(t, \omega) \omega \frac{\partial n(\omega)}{\partial \omega}. \label{eq:heat_1_HO_omega}   
\end{align}
It is an immediate consequence from the definition of energy for the composite system that the first law is satisfied. Indeed $\dot E^{(1)} = \dot F + T \dot S^{(1)}  = \dot W^{(1)}+ \dot Q^{(1)}$ can be easily verified. Obviously,  all reversible changes in the composite system  are first order in the driving rate $\dot \Omega$.

Next we extend our discussion to the variations that occur at a small but finite speed, focusing on the nonequilibrium thermodynamics of the system. Following Ref.\ \citenum{Bruch2016} we adopt a dynamical approach based on the nonequilibrium Green's functions formalism together with the gradient expansion approximation.  As outlined in Appendix  \ref{ap:grad_expan}, this approach yields a nonequilibrium correction to the boson distribution function as experienced by the primary boson, $n(\omega) \to \phi_1(t, \omega)$, that can be obtained from the reduced density matrix of the primary boson.  The result reads  
\begin{align}
\phi_1(t,\omega)=& n(\omega)+\frac{\dot \Omega}{2} A(t, \omega) \frac{\partial}{\partial \omega} n(\omega). \label{eq:noneq_dist_omega}  
\end{align}
  Following Ref.\ \citenum{Bruch2016}, we define nonequilibrium rates in such a way that in the limit of infinitely slow driving we recover the reversible quantities derived above. Nonequilibrium rates will contain higher order corrections in the driving rate $\dot \Omega$ and we will introduce definitions that respect energy balance at each order. In brief, our strategy consists of extending the rates derived for the reversible case by substituting the Boltzmann distribution $n(\omega)$ by the nonequilibrium distribution given by Eq.\ \eqref{eq:noneq_dist_omega}.  Thus, starting from Eq.\ \eqref{eq:revenergy} we postulate the following form for the nonequilibrium energy: 
\begin{align}
  E^{(1)} =& \int \frac{d\omega}{2 \pi} A(t, \omega) \omega \phi_1(t,\omega), \label{eq:HO_noneq_energy}    
\end{align}
such that  $E^{(1)} = E^{(0)} + (\dot \Omega/2) \int d\omega \omega A^2 \partial_\omega n(\omega)$. The definition in Eq.\ \eqref{eq:HO_noneq_energy} is consistent with the rate in Eq.\ \eqref{eq:derE1} up to first order in the modulation rate $\dot \Omega$. Likewise, the nonequilibrium heat and work rates are obtained by extending Eqs.\ \eqref{eq:work_1_HO_omega} and \eqref{eq:heat_1_HO_omega}, that is 
\begin{align}
  \dot W^{(2)} =& \dot \Omega \int \frac{d\omega}{2 \pi}A(t, \omega) \phi_1(t, \omega)
  \nn \\ =& 
 \dot W^{(1)}+\frac{(\dot \Omega )^2}{2} \int \frac{d \omega}{2 \pi} A^2  \frac{\partial}{\partial \omega} n(\omega) \label{eq:work_omega_2} ,  
   \\
\dot Q^{(2)} =& \dot \Omega \int \frac{d\omega}{2 \pi} A(t, \omega) \omega \frac{\partial \phi_1(t, \omega)}{\partial \omega}
\nn \\  =&
\dot Q^{(1)}+\frac{(\dot \Omega)^2}{2} \int \frac{d \omega}{2 \pi} A \omega \frac{\partial}{\partial \omega}\left(  A \frac{\partial n(\omega)}{\partial \omega}\right) . \label{eq:heat_omega_2}   
\end{align}
These definitions are consistent with the energy definition in Eq.\ \eqref{eq:HO_noneq_energy}  for the system, and the identity $\dot E^{(2)} = \dot W^{(2)}+\dot Q^{(2)}$ holds.

  Consider now the entropy production. In studying the driven resonant electron level model it was suggested that the nonequilibrium form for the entropy function can be obtained from its equilibrium form by replacing  the Fermi function by the corresponding nonequilibrium distribution \cite{Bruch2016}. An equivalent assumption would lead to an expression for the entropy given by Eq.\ \eqref{eq:entropy} with $n(\omega)$ replaced by $\phi_1(\omega)$ of Eq.\ \eqref{eq:noneq_dist_omega}.  Such strategy appears to fail in the systems  investigated here. Still, since our main concern are variations in the entropy, we can circumvent the actual definition of a nonequilibrium entropy functional and consider the latter directly. Starting from Eq.\ \eqref{eq:entropy} and the quasistatic evolution derived from the differential  $dS^{(0)} = \partial_\Omega S d \Omega$, we postulate that a local variation in the nonequilibrium entropy functional may be presented in a similar form, provided that $n(\omega)$ is replaced by $\phi_1(t,\omega)$  in $\partial_\Omega S$. This leads to
\begin{align}
\frac{dS}{d t} =& \dot \Omega \frac{\partial S(\phi_1(\omega))}{\partial \Omega}, \label{eq:dotS_omega} . 
\end{align}
assumed correct to second order in $\dot \Omega$, and consequently to the following identity for the rate of entropy change to second order in $\dot \Omega :$ 
\begin{align}
  T \frac{dS^{(2)}}{dt} =& -\frac{\dot \Omega ^2}{2} \int \frac{d\omega}{2 \pi} \left[ A^2 \frac{\partial }{\partial \omega} n(\omega)+A \omega \frac{\partial}{\partial \omega}\left(A \frac{\partial }{\partial \omega} n(\omega) \right) \right].\label{eq:entropy_rate_omega}  
\end{align}
We identify the first term in the integral in Eq.\ \eqref{eq:entropy_rate_omega} with the extra power needed to vary $\Omega$ at a finite rate (as is indeed given by Eq.\ \eqref{eq:work_omega_2}). This term corresponds to the entropy production caused by driving the  system at such finite rate. The second integral in Eq.\ \eqref{eq:entropy_rate_omega} is the second order contribution to the heat transferred to the external bath as follows from Eq.\ \eqref{eq:heat_omega_2}.  We conclude that the present approach to the dynamics and quantum thermodynamics of the slowly driven damped harmonic oscillator brings consistent results in the strong-coupling regime.

\subsection{Driving the coupling strength} \label{sec:HOGamma}

  A different form for time-dependent perturbation appears when we modulate the system-bath coupling strength which is now characterized by the time-dependent parameter $\Gamma(t)$. Again, if the driving rate is slow,  we can assume that the system changes quasistatically and find the retarded Green's function by substitution of  $\Gamma$ by $\Gamma(t)$ in Eq.\ \eqref{eq:GrHO_ss}. As a result we get:
\begin{align}
  G^{r}(t, \omega) =& \frac{1}{\omega - \Omega +i(\Gamma(t)/2)}\label{eq:GrHO_Gamma}.  
\end{align}
Then the spectral density of states is a time-dependent function given by
\begin{align}
  A(t, \omega) =& \frac{\Gamma(t)}{(\omega - \Omega)^2 +(\Gamma(t)/2)^2},  
\end{align}
and the following relation between partial derivatives is satisfied
\begin{align}
  \frac{\partial}{\partial \Gamma} A(t, \omega) =& -\frac{\partial}{\partial \omega} {\rm Re} G^r (t, \omega). \label{eq:relation_gamma}   
\end{align}
 
 The equilibrium thermodynamics of the system is again derived from the canonical potential introduced by Eq.\ \eqref{eq:GrandPot}, as well as from the equilibrium entropy given by Eq.\ \eqref{eq:entropy}. While the driving is different from that considered above, the maximum work principle and the relation between reversible heat and entropy still hold, thus the differential relations $dW = \partial_\Gamma F d \Gamma$ and $dQ = T dS = T \partial_\Gamma S d\Gamma$ remain valid. Therefore, the adiabatic rates of changes in work and heat generated by the reversible driving in the coupling strength can be presented as follows:
\begin{align}
  \dot W^{(1)} =& \dot \Gamma \frac{\partial}{\partial \Gamma}F = \frac{\dot \Gamma}{\Gamma} \int \frac{d \omega}{2 \pi} A(t, \omega)(\omega - \Omega)  n(\omega) \label{eq:work_1_HO_gamma}  
  \\
\dot Q^{(1)} =& T \dot \Gamma \frac{\partial}{\partial \Gamma} S= \frac{\dot \Gamma}{\Gamma} \int \frac{d \omega}{2 \pi} A(t, \omega) (\omega - \Omega) \omega \frac{\partial n(\omega)}{\partial \omega}. \label{eq:heat_1_HO_gamma}  
\end{align}
 As before, the equilibrium relationship $E^{(0)}= F+ T S^{(0)}$ implies that the first law $\dot E^{(1)} = \dot W^{(1)} + \dot Q^{(1)}$ is satisfied to this order.

 \begin{figure}[thb]
   \centering
   \includegraphics[scale=0.72]{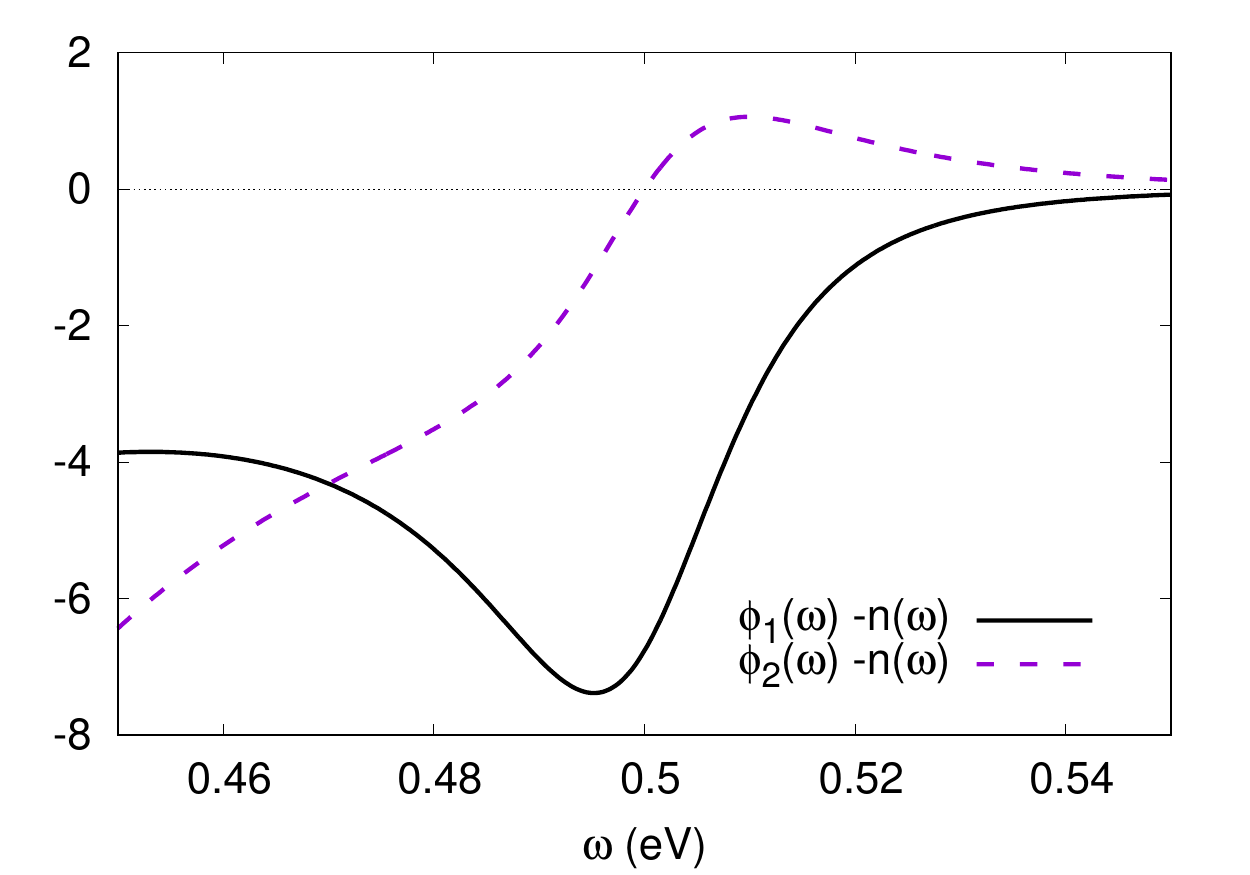}
   \caption{(Color online) Difference between the nonequilibrium distributions for the driven dissipative harmonic oscillator and the Bose-Einstein distribution near the oscillator frequency $\Omega$. The model in consideration has as parameters $\Omega = 0.5$ eV, $\Gamma = 0.03$ eV, $T = 300$ K. In the figure, we plot the difference $\phi_1(\omega)-n(\omega)$ for a linear rate in $\Omega$ of $\dot \Omega = 1$ meV/fs (Solid- black) as well as the difference $\phi_2(\omega)-n(\omega)$ for a linear rate in $\Gamma$ of $\dot \Gamma = 1$ meV/fs (Dashed - purple).}
   \label{fig:dist}
 \end{figure}

 Beyond reversible driving, the nonequilibrium thermodynamics is obtained after identifying the nonequilibrium form for the distribution function, experienced by the primary boson. As detailed in Appendix \ref{ap:grad_expan}, the nonequilibrium Green's functions technique and the gradient expansion approximation provides the functional form for such distribution: 
   \begin{align}
\phi_2(t,\omega)=& n(\omega)-\frac{\dot \Gamma}{2} {\rm Re}G^r \frac{\partial}{\partial \omega} n(\omega).\label{eq:noneq_dist_gamma}   
\end{align}

 The resemblance in the structure of the distributions of Eqs.\ \eqref{eq:noneq_dist_omega} and \eqref{eq:noneq_dist_gamma} is evident, but they behave differently when the frequency $ \omega$ is close to $\Omega$ (see Fig.\ \ref{fig:dist}),  since $A$ and ${\rm Re}G^r$ have different symmetries around the primary boson frequency. Consequently, the dynamical behaviors associated with driving $\Omega$ and $\Gamma$ will be different.

 Repeating the considerations that lead to Eqs.\ \eqref{eq:work_omega_2} and \eqref{eq:heat_omega_2},  we again obtain expression for the rats in which the system exchange work and heat due to $\Gamma$ variations up to order $\dot \Gamma^2$ by replacing $n(\omega)$ by  $\phi_2(\omega)$  in the expressions for the reversible rates Eqs.\ \eqref{eq:work_1_HO_gamma} and \eqref{eq:heat_1_HO_gamma} 
\begin{align}
  \dot W^{(2)} =& \frac{\dot \Gamma}{\Gamma} \int \frac{d \omega}{2 \pi} A(t, \omega)(\omega - \Omega)  \phi_2 (t, \omega)
  \nn \\  =& 
\dot W^{(1)}-\frac{(\dot \Gamma )^2}{2} \int \frac{d \omega}{2 \pi}\left({\rm Re}G^r \right)^2 \frac{\partial}{\partial \omega} n(\omega)\label{eq:work_2_gamma} 
  \\
\dot Q^{(2)} =& \frac{\dot \Gamma}{\Gamma} \int \frac{d \omega}{2 \pi} A(t, \omega) (\omega - \Omega) \omega \frac{\partial \phi_2(t,\omega)}{\partial \omega} 
\nn \\ =&
\dot Q^{(1)}-\frac{(\dot \Gamma)^2}{2} \int \frac{d\omega}{2 \pi} {\rm Re }G^r \omega \frac{\partial}{\partial \omega}\left(  {\rm Re}G^r \frac{\partial n(\omega)}{\partial \omega}\right) \label{eq:heat_2_gamma}  
\end{align}
The time-dependent energy for the composite system is again given by Eq.\ \eqref{eq:HO_noneq_energy}, this time with the nonequilibrium distribution given by Eq.\ \eqref{eq:noneq_dist_gamma}. Consequently, energy conservation (the first law) is established also at the second order in the driving rate $\dot \Gamma$.
 
 Finally, we verify that these rates are consistent with the time derivative of the nonequilibrium entropy. While we do not introduce an explicit expression for this function, we can find a suggestive form for its time derivative to second order in $\dot \Gamma$ by repeating the procedure that lead to Eq.\ \eqref{eq:entropy_rate_omega}, replacing the function $n(\omega)$ in the $\Gamma$-derivative of the entropy functional, $ \partial_\Gamma S(n(\omega))$, by $\phi_2(t,\omega)$, leading to $\dot S = \dot \Gamma \partial_\Gamma S\ (\phi_2)$ correct to second order and hence

 \begin{align}
T \frac{d S^{(2)}}{d t} = &  - \frac{\dot \Gamma^2}{2} \int \frac{d\omega}{2 \pi} \bigg [ \big({\rm Re}G^r \big)^2 \frac{\partial n(\omega)}{\partial \omega}  
  \nn \\    & +
\omega {\rm Re}G^r \frac{\partial}{\partial \omega} \left( {\rm Re} G^r \frac{\partial}{\partial \omega} n(\omega) \right) \bigg] \label{eq:entrop_prod_gamma} .  
 \end{align}

  Here, the first term in the integral 
corresponds to the entropy production (see Eq.\ \eqref{eq:work_2_gamma}) while the second one is the entropy change due to heat transfer (see  Eq.\ \eqref{eq:heat_2_gamma}). Once more, we have found a consistent dynamics as well as thermodynamic description for the damped harmonic oscillator under slow driving.

\subsection{Including effects due to the bath band structure}\label{sec:osc_NWBA}
 In Secs.\ \ref{sec:HO_omega} and \ref{sec:HOGamma} we have neglected the effect of variations in the density of relevant bath modes (modes with $\omega \sim \Omega$) upon variation of $\Omega$. Here we go one step beyond this approximation and consider the situation in which the coupling parameter $\Gamma$ Eq.\ \eqref{eq:gamma_def} varies due to bath band structure. We still assume that $\Gamma$ depends on $\omega$ weakly enough ($\partial \Gamma/ \partial \omega \ll 1$) over the interval of modulation. In this case we can expect that the spectral function $A$ be well described by the Lorentzian 
  \begin{align}
    A(t, \omega) =& \frac{\Gamma(\Omega(t))}{(\omega-\Omega(t))^2+(\Gamma(\Omega(t))/2)^2}, \label{eq:Acolored}
  \end{align}
where we have included the functional dependence of $\Gamma$ on the oscillator's frequency $\Omega$. The spectral function in Eq.\ \eqref{eq:Acolored} satisfies the following identity
\begin{align}
  \frac{\partial}{\partial \Omega}A(t, \omega) =& -\frac{\partial}{\partial \omega}A(t, \omega) - \frac{\partial \Gamma}{\partial \Omega} \frac{\partial}{\partial \omega}{\rm Re}G^r(t, \omega),\label{eq:derAcolored}
\end{align}
which as in previous sections can be used to obtain the rates of change in heat and work due to modulation in $\Omega$. In Eq.\ \eqref{eq:derAcolored} and below, we disregard derivatives of $\Gamma$ with respect to $\omega$ since our considerations allow us to assume that this term is only a function of $\Omega$. The steps involved in the derivation of energy fluxes have been illustrated above: Starting from the canonical potential in Eq.\ \eqref{eq:GrandPot}, this time defined in terms of the spectral function $A$ in Eq.\ \eqref{eq:Acolored}, we obtain equilibrium entropy and energy functionals in the corresponding forms given by Eqs.\ \eqref{eq:entropy} and \eqref{eq:revenergy} (with $A$ given by Eq.\ \eqref{eq:Acolored}). The reversible work $\dot W^{(1)}$ and heat rates $\dot Q^{(1)}$ are derived from the maximum work principle and the fact that quasistatic heat due to infinitesimal transformation is proportional to the infinitesimal change in the entropy of the system.  As a consequence of the relation \eqref{eq:derAcolored}, we find that the heat and work rates can each be written in terms of two contributions: direct modulation in $\Omega$ as well as a correction term, proportional to $\partial \Gamma/ \partial \Omega$, originating from the indirect modulation in $\Gamma$. The explicit form of the reversible rates are proportional to $\dot \Omega$ and given by     
\begin{align}
  \dot W^{(1)} =& \dot \Omega \int \frac{d\omega}{2 \pi} A(t, \omega) n(\omega) \notag\\
&+\frac{\dot \Omega }{\Gamma} \frac{ \partial \Gamma}{\partial \Omega} \int \frac{d \omega}{2 \pi} A(t, \omega)(\omega - \Omega)  n(\omega)\\
  \dot Q^{(1)} =& \dot \Omega \int \frac{d \omega}{2 \pi} A(t, \omega) \omega \frac{\partial n(\omega)}{\partial \omega} \notag\\
&+ \frac{\dot \Omega }{\Gamma} \frac{ \partial \Gamma}{\partial \Omega} \int \frac{d \omega}{2 \pi} A(t, \omega) (\omega - \Omega) \omega \frac{\partial n(\omega)}{\partial \omega}.
\end{align}

Beyond quasistatic dynamics and utilizing the results in Appendix \ref{ap:grad_expan}, we find the nonequilibrium distribution function $\tilde \phi(t, \omega)$ valid to first order in $\dot \Omega$  
\begin{align}
  \tilde \phi(t, \omega) =& n(\omega)+ \dot \Omega(t) \left(  A(t, \omega) - \frac{1}{2} \frac{ \partial \Gamma}{\partial \Omega} {\rm Re}G^r \right) \frac{\partial}{\partial \omega}n(\omega).
\end{align}
Repeating the considerations that lead to Eqs.\ \eqref{eq:work_omega_2} and \eqref{eq:heat_omega_2}, we once more obtain expression for $\dot W^{(2)}$ and $\dot Q^{(2)}$. We notice that the nonequilibrium rates up to second order in the driving rate $\dot \Omega$ include corrections due to the bath structure that are proportional to $(\partial \Gamma/ \partial \Omega)^2$:
\begin{align}
  \dot W^{(2)} =& \dot W^{(1)}+\frac{(\dot \Omega )^2}{2} \int \frac{d \omega}{2 \pi} A^2  \frac{\partial}{\partial \omega} n(\omega) \notag\\
&-\frac{(\dot \Omega )^2}{2}\left(\frac{\partial \Gamma }{\partial \Omega}\right)^2 \int \frac{d \omega}{2 \pi}\left({\rm Re}G^r \right)^2 \frac{\partial}{\partial \omega} n(\omega) \label{eq:work_2_NWBA}\\
 \dot Q^{(2)} =& \dot Q^{(1)}+\frac{(\dot \Omega)^2}{2} \int \frac{d \omega}{2 \pi} A \omega \frac{\partial}{\partial \omega}\left(  A \frac{\partial n(\omega)}{\partial \omega}\right) \notag\\
&-\frac{(\dot \Omega )^2}{2}\left(\frac{\partial \Gamma }{\partial \Omega}\right)^2 \int \frac{d\omega}{2 \pi} {\rm Re }G^r \omega \frac{\partial}{\partial \omega}\left(  {\rm Re}G^r \frac{\partial n(\omega)}{\partial \omega}\right)\label{eq:heat_2_NWBA}
\end{align}
Finally, we remark that the entropy rate 
\begin{align}
  T \frac{dS^{(2)}}{dt} =& -\frac{\dot \Omega ^2}{2} \int \frac{d\omega}{2 \pi} \left[ A^2 \frac{\partial }{\partial \omega} n(\omega)+A \omega \frac{\partial}{\partial \omega}\left(A \frac{\partial }{\partial \omega} n(\omega) \right) \right]\notag\\
&- \frac{\dot \Omega^2}{2} \left(\frac{\partial \Gamma}{\partial \Omega}\right)^2 \int \frac{d\omega}{2 \pi} \bigg [ \big({\rm Re}G^r \big)^2 \frac{\partial n(\omega)}{\partial \omega}  
  \notag \\    & \hspace{2cm}+
\omega {\rm Re}G^r \frac{\partial}{\partial \omega} \left( {\rm Re} G^r \frac{\partial}{\partial \omega} n(\omega) \right) \bigg], 
\end{align}
also includes correction terms proportional to  $(\partial \Gamma/ \partial \Omega)^2$ and is consistent with the rates obtained in Eqs.\ \eqref{eq:work_2_NWBA} and \eqref{eq:heat_2_NWBA}.


\section{The Damped Two-Level System}\label{sec:TLS}

 In this section, we consider a two-level molecule strongly coupled with a thermal bath represented, as before, by a continuum of harmonic modes. We will again disregard changes in the local bath band structure by adopting a wide band approximation. The methods introduced in Sec.\ \ref{sec:osc_NWBA} can be implemented here if one needs to account for the effect of such structural change. In the Hilbert space of the molecule, each level is represented by a ket $| i \rangle$, with $i \in \{ 1,2 \}$. The Hamiltonian for the composite system is the sum of the free Hamiltonian for the molecule $\hat H_{\rm TLS}$, the harmonic bath Hamiltonian $\hat H_{B}$ and the coupling $V$:
\begin{align}
 & \hat H = \hat H_{\rm TLS} + \hat H _{\rm B} + \hat V, \label{eq:Hamil_TLS}  
  \\ &
\hat H_{\rm TLS} = \omega_L \hat \sigma^z ,  \label{eq:HsysTLS}
\\ &
\hat H_{\rm B} = \sum_k \omega_k \hat b_k ^\dagger \hat b_k ,
 \\ &
\hat V = i\frac{1}{2} \sum_k \left( u_k \hat \sigma^{+} \hat b_k - u_k^* \hat \sigma^{-} \hat b_k^{\dagger}\right) \label{eq:HintTLS} 
\end{align}
where $\hat \sigma^{z} = (1/2) \left(| 2 \rangle \langle 2 | - | 1 \rangle \langle 1 | \right),\  \hat \sigma^{+} = |2 \rangle \langle 1|$ and $ \hat \sigma^{-} = |1 \rangle \langle 2|$.  Here, $ \omega_L$ is the spacing between level energies, $ \omega_k $ are the frequencies of the bath modes and $ u_k $ are the molecule-bath coupling elements.  A complete thermodynamic description at equilibrium can be obtained from the free energy -- the canonical potential for the two-level-system-bath composite system. The partition function and the free energy for this model are calculated in Appendix \ref{ap:grand_pot_TLS} from an approximate description of the energy spectrum of the two-level system interacting with a finite but large bath. In the derivation we assume that the energy spacing between consecutive modes in the bath is small and we take the limit of infinitesimal spacing. The result reads:
\begin{align}
  F = & \frac{1}{\beta} \int \frac{d \omega}{2 \pi}\mathcal{A}(\omega) \ln (1 - e^{-\beta \omega})- \frac{1}{2}\sqrt{\omega_L^2 + 4 \eta} \label{eq:GCP_TLS}   
\end{align}
where
\begin{align}
\eta =& \lim_{N \to \infty} (1/4 N) \sum_{k=1}^N |u_k|^2,  \label{eq:eta_tls}
\end{align}
 and $\mathcal{A}(\omega)$ represents the spectral density.  In standard models for thermal baths, $\sum_k |u_k|^2$ is constant and $\eta \to 0$ as $N \to \infty$. Again the equilibrium entropy functional is obtained by differentiation of the canonical potential in Eq.\ \eqref{eq:GCP_TLS} with respect to the absolute temperature $T.$  As a result, we arrive at the following expression: 
\begin{align}
  S^{(0)}=& -k_B \int \frac{d \omega}{2 \pi} \mathcal{A}(\omega) \Big[ n(\omega) \ln (n(\omega)) 
  \notag \\
&\hspace{2.5cm} -(1+ n(\omega))\ln (1 + n(\omega))\Big].  \label{eq:entropy_tls}  
\end{align}
 An approximate expression for the spectral density $\mathcal{A}(\omega)$ is found  using the NEGF technique in Appendix \ref{ap:Gr_TLS}. We get:
\begin{align}   
  \mathcal{A}(\omega) =&  \frac{\Gamma \mathcal{S}^2}{(\omega - \omega_L)^2 + (\Gamma \mathcal{S}/ 2)^2}\label{eq:DOS_tls}.
\end{align}    
In this expression, $\mathcal{S} = - 2 \langle \hat \sigma^z \rangle$  is the difference in population between the levels. The approximation employed to obtain Eq.\ \eqref{eq:DOS_tls} assumes a factorization of a higher order correlation function in terms of lower order ones, providing a simple solution to the associated Dyson equation (see Eq.\ \eqref{eq:ap_approx_Gr}). We notice that in the absence of population inversion $\mathcal{S}$ is positive. If the change in $\omega_L$ due to driving is small relative to $\omega_L$ itself, we may disregard the dependence of $\mathcal{S}$ on $ \omega_L. $ In this case  the spectral function $\mathcal{A}$ defined by Eq.\ \eqref{eq:DOS_tls} satisfies the equation
\begin{align}
  \frac{\partial}{\partial \omega} \mathcal{A}(\omega) =& - \frac{\partial}{\partial \omega_L} \mathcal{A}(\omega) .\label{eq:der_tls}   
\end{align}
 This property of $\mathcal{A} $ is used in the following computations of work and heat rates. 
 
 The equilibrium energy functional $E^{(0)}=F+TS^{(0)}$ can be determined from Eqs.\ \eqref{eq:GCP_TLS} and \eqref{eq:entropy_tls} and is given explicitly by the expression
 \begin{align}
   E^{(0)} =& \int \frac{d\omega}{2 \pi} \mathcal{A}(\omega) \omega n (\omega) - \frac{1}{2}\sqrt{\omega_L^2+ 4 \eta}. \label{eq:energy_tls}   
 \end{align}
        Next, we introduce the quasistatic work and heat rates, utilizing as in the previous section the maximum work principle and the relation between entropy change and reversible heat. This leads to
\begin{align}
\dot W^{(1)} = & \dot \omega_L \frac{\partial}{\partial \omega_L} F      
\nn \\ =& 
\dot \omega_L \int \frac{d \omega}{2 \pi} \mathcal{A}(\omega) n(\omega) - \frac{\dot \omega_L}{2}  \frac{\omega_L}{\sqrt{\omega_L^2+ 4 \eta}} \label{eq:work_1_tls}
  \\
\dot Q^{(1)} =& \frac{k_B}{\beta} \dot \omega_L \frac{\partial}{\partial \omega_L} S 
 =   \dot \omega_L \int \frac{d\omega}{2 \pi} \mathcal{A}(\omega) \omega \frac{\partial n (\omega)}{\partial \omega_L} \label{eq:heat_1_tls}.  
\end{align}
 The definition of the equilibrium energy $E^{(0)}$ and the fact that the  quasistatic energy variation is given by $\dot E^{(1)} = \dot \omega_L \partial E^{(0)}/\partial \omega_L$ imply that energy balance (the first law) holds for the rates derived in Eqs.\ \eqref{eq:work_1_tls} and \eqref{eq:heat_1_tls}, that is $\dot E^{(1)} = \dot W^{(1)}+\dot Q^{(1)}$. 

 It is interesting to compare the quasistatic evolutions of this system and the damped harmonic oscillator considered in Sec.\ \ref{sec:HO_omega}. In the limit of large separation between levels, $\mathcal{S} \to 1.$ Then Eqs.\ \eqref{eq:heat_1_HO_omega} and \eqref{eq:heat_1_tls}   yield identical expressions for reversible heat rates provided that $\Omega $ is identified with $ \omega_L . $ The expressions for the reversible work flux in Eqs.\ \eqref{eq:work_1_HO_omega} and \eqref{eq:work_1_tls} appear different, however this difference (which is also reflected by the second term in Eq.\ \eqref{eq:work_1_tls},  just reflects the fact the the ground state of the two-level system was chosen to be $-\omega_L/2$ (Note that $\eta$ in Eq.\ \eqref{eq:eta_tls} vanishes if $u_k$ is constant independent of the number of modes taken to model the bath).
 
 As before nonequilibrium effects appear in the next order (2) in $\dot \omega_L$ and explicit expressions can be  derived following the procedure used previously. First, we find the nonequilibrium distribution function ( see Appendix \ref{ap:noneq_dis_TLS}).
\begin{align}
  \phi_3(t, \omega) =& n(\omega)+ \frac{\dot \omega_L}{2}\mathcal{S}^{-1}\mathcal{A}(t, \omega)\frac{\partial}{\partial \omega} n(\omega).  \label{eq:dist_tls}   
\end{align}
 Then we employ this function to compute the work and heat nonequilibrium rates. For this purpose, we replace the Bose-Einstein distribution functions in the expressions 	\eqref{eq:work_1_tls} and \eqref{eq:heat_1_tls},  by $ \phi_3(t,\omega).$ The resulting nonequilibrium rates equal
\begin{align}
  \dot W^{(2)} =& \dot \omega_L \int \frac{d \omega}{2 \pi} \mathcal{A}(\omega) \phi_3(t, \omega) - \dot \omega_L  \frac{\omega_L}{2\sqrt{\omega_L^2+ 4 \eta}}
\nn \\= &
\dot W^{(1)}+ \frac{(\dot \omega_L)^2}{2} \mathcal{S}^{-1}\int \frac{d \omega}{2 \pi} \mathcal{A}^2 \frac{\partial n(\omega)}{\partial \omega} \label{eq:work_2_tls} 
 \\
\dot Q^{(2)}=&  \dot \omega_L \int \frac{d\omega}{2 \pi} \mathcal{A}(\omega) \omega \frac{\partial \phi_3 (\omega)}{\partial \omega_L}
\nn \\ =& 
\dot Q^{(1)}+ \frac{(\dot \omega_L)^2}{2}\mathcal{S}^{-1} 
\int \frac{d\omega}{2 \pi} \mathcal{A} \omega \frac{\partial}{\partial \omega}\left[ \mathcal{A} \frac{\partial}{\partial \omega}n(\omega) \right].\label{eq:heat_2_tls}  
\end{align}
Also, making the same replacement $ [n(\omega) \to \phi_3(\omega)] $ in the expression \eqref{eq:energy_tls} for the energy functional we can verify  that energy balance holds at second order in $\dot \omega_L$ for the rates given by  Eqs.\ \eqref{eq:work_2_tls} and \eqref{eq:heat_2_tls}.

Similarly, the second order contributions to the total entropy rate $\dot S^{(2)}$, calculated from the differential $dS = (\partial_{\omega_L} S) d \omega_L$,  permit a full identification of the entropy production term. Indeed:
 \begin{align}
   T \frac{d S}{dt} =& -\frac{\dot \omega_L}{2} \mathcal{S}^{-1} \int \frac{d\omega}{2 \pi} \left[ \mathcal{A}^2 \frac{\partial n(\omega)}{\partial \omega} + \omega \mathcal{A}  \frac{\partial}{\partial \omega}\left( \mathcal{A}\frac{\partial}{\partial \omega} n(\omega) \right) \right],\label{eq:entropy_rate_tls} 
 \end{align}
  Here, the first integral on the right hand side of Eq.\ \eqref{eq:entropy_rate_tls} corresponds to the rate of heat dissipated as entropy production which is already identified by Eq.\ \eqref{eq:work_2_tls} as the nonequilibrium work rate $\dot W^{(2)}$, while the second integral is the heat flux determined by Eq.\ \eqref{eq:heat_2_tls}.  Thus we have achieved a complete and consistent dynamic as well as thermodynamic representation of the damped two-level system under reversible and slow driving of the energy gap $\omega_L$.
 
Finally note that (as expected) also the second order terms are the same as the damped harmonic oscilator in the limit $\mathcal{S} \to 1$.


\section{Friction}\label{sec:friction}

 Dissipation in a nanoscale engine due to its interactions with the environment could be introduced in the equations of motion for the system describing time evolution of a physical coordinate by adding a phenomenological friction term.  The analytic form for  dissipative terms which may be ascribed to friction can be singled out from the detailed quantum mechanical description of the dynamics of a particular open system. As known, friction is closely related to the power dissipated in the system. It strongly depends on the system's speed. When the motion is infinity slow friction approaches zero, and it increases as the system is speeding up\cite{rezek2006irreversible}. In Eqs.\ \eqref{eq:work_omega_2}, \eqref{eq:work_2_gamma} and \eqref{eq:work_2_tls} we have identified the power dissipated under finite speed in the damped harmonic oscillator and in the dissipative two-level molecule subject to various drivings. In order to define a friction coefficient for each case, we have to associate time perturbations with  changes in certain external coordinates. Below we use an example to show how these relationships may be established.

 In a recent experimental work (Ref.\ \citenum{Rossnagel2016}), an Otto engine was realized with a single trapped ion in a linear Paul trap with a funnel-shaped electrode geometry. The radial trap frequency $\omega_{x,y}$ was observed to be descreasing in the axial $z$-direction as  
 \begin{align}
   \omega_{x,y} = {\omega_o}\bigg /{\ds\left(1 + \frac{z}{r_o} \tan \theta \right)^2}.   
 \end{align}
This result suggests that the approximation, $\omega_{x,y} = \omega_o (1 - 2 z \tan \theta/ r_o)$ may be employed for small $\theta. $ Thus, a displacement along the $z$ axis in the trap induces a change of frequency $\dot \omega_{x,y} = - 2 \tan \theta \dot z$. This demonstrates that a linear relation between the characteristic frequency of an atomic oscillator and physical displacement is feasible. Thus,  for the damped harmonic oscillator, with the driven  $\Omega$ (Sec.\ \ref{sec:HO_omega}), we can assume that $\dot \Omega =  M_1  \dot z .$ 

 We may generalize this relationship and  apply it to our model. The dissipated power $\dot W^{(2)}$ is caused by a friction force $F_1$ acting on the external coordinate $z$ according to $ \dot W^{(2)} = - F_1 \dot z$, with $F_1 = - \gamma_1 \dot z$. Then Eq.\ \eqref{eq:work_omega_2} leads to the following form for the friction coefficient $\gamma_1:$
 \begin{align}
\gamma_1 =& - \frac{M_1^2}{2} \int \frac{d \omega}{2 \pi} A^2  \frac{\partial}{\partial \omega} n(\omega).  
 \end{align}
Similarly if the rate of changes $\dot \Gamma$ in Eq.\ \eqref{eq:work_2_gamma} and $\dot \omega_L$ in Eq.\ \eqref{eq:work_2_tls}  could be related to some coordinate $z$ via $\dot \Gamma = M_2 z$ and $\dot \omega_L = M_3 z$, then the corresponding friction coefficients for motions along these coordinates would be 
\begin{align}
\gamma_2 =& - \frac{M_2^2}{2} \int \frac{d \omega}{2 \pi}\left({\rm Re}G^r \right)^2 \frac{\partial}{\partial \omega} n(\omega),\label{eq:fric2}  
\end{align}

\begin{align}
\gamma_3 =& - \frac{M_3^2}{2} \mathcal{S}^{-1}\int \frac{d \omega}{2 \pi} \mathcal{A}^2 \frac{\partial n(\omega)}{\partial \omega}.\label{eq:fric3}  
\end{align}


\section{Conclusions}\label{sec:conclusions}

 We have presented a systematic description of the dynamics as well as the thermodynamics for a harmonic oscillator and a two-level system coupled to a harmonic bath, both subject to slow driving rates. Our approach is an extension of that one introduced in Ref.\ \citenum{Bruch2016}. The effects of driving are studied within the nonequilibrium Green's functions formalism and the gradient expansion method. Our results are consistent with the first and second laws of thermodynamics, yielding explicit expressions for the work, heat and entropy productions associated with the driving process, valid for system bath interactions of arbitrary strengths. Similar to Ref.\ \citenum{Bruch2016} (see also Ref.\  \citenum{ochoa2016energy}) we could identify, within the models studied, and effective system Hamiltonian that accounts for system properties by including half the system-bath interaction.  Unlike Ref. \citenum{Bruch2016}, a suggestive expression for the entropy production rate is obtained without the need to define the total entropy.

 The formalism introduced in the present work can provide a guideline
for future thermodynamic treatments of strongly coupled quantum nanoscale systems, and can be directly applied to currently explored experimental setups such as realized optomechanical heat engine\cite{zhang2014quantum,dechant2015all} or an approach of a molecule to a metal surface.

\appendix

\section{Retarded Green function for the damped harmonic oscillator $G^r$}\label{ap:GrHO}

Here we derive Eq.\ \eqref{eq:GrHO_ss}. From the Hamiltonian given by Eq.\ \eqref{eq:Hamil} we find that the Heisenberg Equations of motion for $\hat a$, and $\hat b_m$ are
\begin{align}
  i \frac{d}{dt}\hat a(t) = \Omega \hat a + \sum_m u _m \hat b_m \label{eq:EOMa}    
  \\
i\frac{d}{dt} \hat b_m(t) = \omega_m \hat b_m +u_m \hat a .\label{eq:EOMbm}   
\end{align}
Next, we derive the equation of motion (EOM) for the Green's function defined in Eq.\ \eqref{eq:GreenF}, in the Keldysh contour to later find its retarded expression in frequency space. Indeed, utilizing Eq.\ \eqref{eq:EOMa} we get
\begin{align}
  i \frac{d}{d\tau_1}G(\tau_1,\tau_2) = &  \delta(\tau_1, \tau_2) + \Omega G(\tau_1, \tau_2)
	\nn \\ &+ \sum u_m G_{ m\, a}(\tau_1, \tau_2) , \label{eq:EOMg}   
\end{align}
with $   G_{ m\, a}(\tau_1, \tau_2) = -i \langle \hat b_m(\tau_1) \hat  a^\dagger(\tau_2) \rangle$. Now we find the EOM for $G_{m \, a}(\tau_1, \tau_2)$ utilizing Eq.\ \eqref{eq:EOMbm}, that is,
\begin{align}
\left(i \frac{d}{d\tau_1} - \omega_m \right) G_{m \, a}(\tau_1, \tau_2) = & u_m G(\tau_1, \tau_2). \label{eq:EOMGma}      
\end{align}
For $g_m (\tau_1, \tau_2) = -i \langle \hat b_m(\tau_1) \hat b_m^\dagger(\tau_2) \rangle$, the Green's function that solves the Dyson equation for a free boson (null self-energy), we verify that the identity 
\begin{align}
  \left(i \frac{d}{d\tau_1} - \omega_m \right) g_m(\tau_1, \tau_2) = & \delta(\tau_1, \tau_2) \label{eq:nB}  
\end{align}
holds. The result described by  Eq.\ \eqref{eq:nB} permits us to solve Eq.\ \eqref{eq:EOMGma}:
\begin{align}
G_{m \, a}(\tau_1, \tau_2) = & u_m \int d\tau_3 g_m(\tau_1,\tau_3 ) G(\tau_3, \tau_2).\label{eq:gma}      
\end{align}
Substituting Eq.\ \eqref{eq:gma} into \eqref{eq:EOMg} we obtain
\begin{align}
 i \frac{d}{d\tau_1}G(\tau_1,\tau_2) =& \delta(\tau_1, \tau_2) + \Omega G(\tau_1, \tau_2)
	\nn\\
& + \sum |u_m|^2 \int d\tau_3\, g_m(\tau_1, \tau_3) G(\tau_3, \tau_2) .   
\end{align}
We now project onto the real line to derive the retarded form $G^{r}(t_1, t_2)$ of the Green's function using Langreth rules. Then, we define new variables $s= t_1- t_2$ and $t=(t_1+t_2)/2$ such that,
\begin{align}
   i \left( \frac{d}{ds}+\frac{1}{2}\frac{d}{dt}\right)G^r(t,s) =& \delta(s) + \left( \Omega  -i \frac{\Gamma}{2}\right) G^r(t, s),\label{eq:EOMst}   
\end{align}
where we have adopted the wide-band limit for the last term in Eq.\ \eqref{eq:EOMst}. We calculate the Fourier transform with respect to $s$ in Eq.\ \eqref{eq:EOMst} to get
\begin{align}
G^r(t, \omega) =& \left(1-\frac{i}{2}\frac{d}{dt}G^r(t,\omega)\right) \left(\frac{1}{\omega - \Omega +i(\Gamma/2)}\right) \label{eq:GrTomega}  . 
\end{align}
Thus the zeroth order approximation for $G^r(t,\omega)$, corresponding to the adiabatic limit, is obtained by disregarding the term involving the derivative with respect to $t$ in the right hand side of Eq. \eqref{eq:GrTomega}.  The result is given in Eq.\ \eqref{eq:GrHO_ss}.

\section{Lower cutoff for the canonical potential}\label{ap:Cutoff}

We introduce a cutoff frequency $\omega_o = 1/n$ such that
\begin{align}
  F(\Omega, \Gamma) =& \int_0^\infty \frac{d\omega}{2 \pi} A(\omega) \ln (1 - e^{-\beta \omega})     
	\nn \\  =& 
\int_{\omega_o}^\infty \frac{d\omega}{2 \pi} A(\omega) \ln (1 - e^{-\beta \omega})
 \nn \\  &\hspace{0.5cm} + 
\int_0^{\omega_o} \frac{d\omega}{2 \pi} A(\omega) \ln (1 - e^{-\beta \omega}),
\end{align}    
 We estimate $\frac{\partial}{\partial \Gamma} F(\Omega, \Gamma) $ to show that the terms below the lower cutoff do not contribute to the rates  $\dot\Gamma$. In the region $(0, \omega_o)$ we approximate $\ln [1 - e^{-\beta \omega}] \approx \ln (\beta \omega)$. Then:
\begin{align}
  & \Big| \frac{\partial}{\partial \Gamma} \int_0^{\omega_o} \frac{d\omega}{2 \pi} A(\omega)  \ln (1 - e^{-\beta \omega}) \Big| 
	 \nn\\
\leq & \int_0^{\omega_o} \frac{d \omega}{2 \pi} \frac{\Omega^2}{\Gamma} A(\omega) |\ln (\beta \omega)|
\nn\\
= & \lim_{n \to \infty} \int_{1/(n+1)}^{\omega_o} \frac{d \omega}{2 \pi} \frac{\Omega^2}{\Gamma} A(\omega) |\ln (\beta \omega)|
\nn \\ 
\leq & - \frac{\Omega^2}{\Gamma}  \lim_{n \to \infty} \ln \left(\frac{\beta}{n+1} \right) \int_{1/(n+1)}^{\omega_o} \frac{d \omega}{2 \pi}  \frac{1}{(\omega - \Omega)^2}
\nn\\
\leq & - \frac{1}{\Gamma} \lim_{n \to \infty} \ln \left(\frac{\beta}{n+1} \right) \left(\frac{1}{n} - \frac{1}{n+1} \right)   \to 0 .            
\end{align}

\section{Effective Hamiltonian for the Extended Harmonic oscillator}\label{ap:Elambda}
In this section we calculate the partial contributions to the total energy of the dissipative harmonic oscillator utilizing the method in Ref.\  \citenum{ochoa2016energy}. In brief we introduce rescaling parameters $(\lambda_S, \lambda_B, \lambda_V)$ in the Hamiltonian in Eq.\ \eqref{eq:Hamil} such that 
\begin{align}
\hat H (\lambda_S, \lambda_B, \lambda_V)=& \lambda_S \hat H_S+\lambda_B \hat H_B+\lambda_V \hat V. \label{eq:Hlambda}  
\end{align}
This rescaling transfers to the spectral function $A$ as well as to the canonical potential according to
\begin{align}
A(\lambda_S, \lambda_B, \lambda_V) =& \frac{\lambda_B^{-1} \lambda_V^2 \Gamma}{(\omega - \lambda_S \Omega)^2+(\lambda_B^{-1} \lambda_V^2 \Gamma)^2}\\
\Omega(\lambda_S, \lambda_B, \lambda_V) =& \frac{1}{\beta} \int A(\lambda_S, \lambda_B, \lambda_V) \ln (1 - e^{-\beta \omega}).
\end{align}
With these definitions we can show that 
\begin{align}
  \frac{\partial}{\partial \lambda_S} A(\lambda_S,1, 1) &=- \Omega \frac{\partial}{\partial \omega} A(\lambda_S,1, 1)\\
  \frac{\partial}{\partial \lambda_B} A(1, \lambda_B 1) &=\lambda_B^{-2}\Gamma \frac{\partial}{\partial \omega} {\rm Re}G^r(1,\lambda_B, 1)\\
  \frac{\partial}{\partial \lambda_V} A(1, 1,\lambda_V) &=- 2\lambda_V\Gamma \frac{\partial}{\partial \omega} {\rm Re}G^r(1,1, \lambda_V)
\end{align}
as well as
\begin{align}
  \langle \hat H_S \rangle =& \Omega \int \frac{d\omega}{2 \pi} A(\omega) n(\omega) \label{eq:Hs_ap}\\
  \langle \hat H_B \rangle =& -\int \frac{d\omega}{2 \pi}(\omega - \Omega) A(\omega) n(\omega)\\
  \langle \hat V \rangle =& 2\int \frac{d\omega}{2 \pi}(\omega - \Omega) A(\omega) n(\omega)\label{eq:V_ap}.
\end{align}
Equations \eqref{eq:Hs_ap} - \eqref{eq:V_ap} follow from the identity 
\begin{align}
  \langle \hat H_i \rangle = -\beta \frac{\partial}{\partial \lambda_i} \Omega(\lambda_i) 
\end{align}

\section{Nonequilibrium distribution functions}\label{ap:grad_expan}

 Starting from the definition in Eq.\ \eqref{eq:GreenF}, we can  implement the gradient expansion and keep only the terms up to first order in energy and time derivatives. We then obtain
\begin{align}
 G^< (t, \omega) 
	=& G^r(t, \omega ) \Sigma^< (t, \omega) G^a(t, \omega)
	\notag\\  &+ 
\frac{i}{2}\Big[ G^r(t, \omega) \frac{\partial G^a(t, \omega)}{\partial t}
\notag\\ & -
\frac{\partial G^r(t, \omega)}{\partial t} G^a(t, \omega) \Big] \frac{\partial \Sigma^<(\omega)}{\partial \omega}\label{eq:grad_ex_ap} .  
\end{align}
 Since 
 \begin{align}
   \frac{\partial G^r}{\partial t} = \dot \Omega (G^r)^2 ,
	&&
	\frac{\partial G^a}{\partial t} = \dot \Omega (G^a)^2  , 
\end{align}
\begin{align}
   G^r G^a =& \frac{A(t, \omega)}{\Gamma},    
 \end{align}
we get:
\begin{align}
  i G^<(t,\omega) =& A n(\omega) + \frac{\dot \Omega}{2}A^2 \frac{\partial}{\partial \omega} n(\omega),     
\end{align}
where $n(\omega)$ is the Bose-Einstein distribution function. We define the nonequilibrium distribution function $\phi_1 (t, \omega)$ by the expression
\begin{align}
  i G^r(t, \omega) =& A(t, \omega) \phi_1 (t,\omega)  . 
\end{align}
Consequently,  $\phi_1 (t, \omega)$ should be given by  Eq.\ \eqref{eq:noneq_dist_omega}. 

 In Sec. \ref{sec:HOGamma} we have studied the quantum thermodynamics when driving affects the coupling strength. In this case and starting from  Eq.\ \eqref{eq:GrHO_Gamma} we have:
 \begin{align}
    \frac{\partial G^r}{\partial t} = -\frac{i}{2}\dot \Gamma (G^r)^2 ,
		&&
	\frac{\partial G^a}{\partial t} = \frac{i}{2} \dot \Gamma (G^a)^2,    
 \end{align}
which after substitution in Eq.\ \eqref{eq:grad_ex_ap} lead to
\begin{align}
  i G^<(t,\omega) =& A n(\omega) -\frac{\dot \Gamma}{2}A {\rm Re}G^r \frac{\partial}{\partial \omega} n(\omega)    
\end{align}
From this expression, we obtain the result for $\phi_2(t,\omega) $ given by Eq.\ \eqref{eq:noneq_dist_gamma}. 

\section{Potential for the Damped two-level system}\label{ap:grand_pot_TLS}

Here we derive the expression for the canonical potential for the dissipative two-level system discussed in Sec.\ \ref{sec:TLS}. We start by studying the Hamiltonian and the energy spectrum of a two-level system coupled to a finite-bath with $N$ noninteracting bosons.  We assume that the frequency of boson mode $k$ in the bath is given by $\omega_k = k \Delta \omega $ ($\Delta \omega$ is the inverse density of modes, assumed constant), with $k \in \{ 0, \dots, N\} $,  $\Delta \omega = \omega_{\rm max}/ N$, and $\omega_{\rm max}$ is an upper frequency cutoff defining the bandwidth of the bath. Moreover, for each mode $k$ we consider a finite number of phonons $n_k$. Thus the bath is characterized by the set of pararameters $\{ N, \Delta \omega, \{n_k\}\}$. System-bath coupling is defined by the Hamiltonian in Eq.\ \eqref{eq:HintTLS}, which assumes the rotating phase approximation.   A basis for the composite system  (TLS + finite bath) is obtained from the tensor product between the energy eigenbasis for the two level system and the diagonal basis for the noninteracting bath:  denoting the two-level system eigenvectors by  $| l \rangle $,  $l \in \{1,2\}$, the basis for the composite state is $| l, \{n_k | 1\leq k \leq N\} \rangle = | l \rangle \otimes | n_1\rangle \otimes \dots \otimes | n_N\rangle $.  In this basis and as a consequence of the interaction Hamiltonian in Eq.\ \eqref{eq:HintTLS}, we find that 
\begin{align}
  \langle 1, n_1,\dots,n_k+1,\dots,n_N | \hat V | 2, n_1,\dots,n_k,\dots,n_N \rangle =& -\frac{i}{2} u_k \label{eq:MatElemOff}
\end{align}
for all $1\leq k\leq N$, and also
\begin{align}
   \langle l, \{ n_k \} | \hat H_{\rm TLS} +\hat H_B | l, \{ n_k \} \rangle =& \frac{(-1)^l}{2}\omega_L + \sum_{k=1}^N \omega_k n_k\label{eq:MatElemDiag}.
\end{align}
Let $\varepsilon_B = \sum_{k=1}^N \omega_k n_k$ and $s=2+\sum_k n_k$. Equations \eqref{eq:MatElemOff} and \eqref{eq:MatElemDiag} indicate that the Hamiltonian acts on the state vector $|l , \{n_k \} \rangle$ by preserving the total number $s$. In particular, for a system in the initial state $| 2, \{ n_k \} \rangle$ allowed transitions couple relaxations at the two-level system ($2 \to 1$) with excitations in a single mode in the bath ($n_k \to n_k +1$ for some $k$). Thus, in the subspace generated by the family of kets
\begin{align}
\big\{& |2, \{n_k\} \rangle, \notag\\
&|1,n_1+1,\{n_k, k\neq1\} \rangle, \notag\\
&\dots, |1,\{n_k, k< j\}, n_j+1,\{n_k, k> j\} \rangle, \notag\\
& \dots, |1,\{n_k, k < N\}, n_N+1 \rangle \big\},   \label{eq:nksubbasis}
\end{align}
we find a matrix representation for the Hamiltonian Eq.\ \eqref{eq:Hamil_TLS}, in terms of matrices $A$ and $B$

\begin{widetext}
\begin{align}
A=& \begin{pmatrix}
    -\frac{\omega_L}{2} + \varepsilon_B&0&0&0\\
    0& -\frac{\omega_L}{2} + \omega_1+\varepsilon_B&0&\dots&0\\
    0& 0&-\frac{\omega_L}{2} +\omega_2+ \varepsilon_B&\dots&0\\
    \vdots& \vdots&\vdots&\ddots&\vdots\\
    0&0&0&\dots&-\frac{\omega_L}{2} + \omega_N+\varepsilon_B\\
  \end{pmatrix} \\
 B=& \begin{pmatrix}
    \omega_L& -\frac{i}{2}u_1& -\frac{i}{2}u_2&\dots&-\frac{i}{2}u_N\\
    \frac{i}{2}u_1&0&0&\dots&0\\
    \frac{i}{2}u_2&0&0&\dots&0\\
    \vdots& \vdots&0&\ddots&\vdots\\
    \frac{i}{2}u_N&0&0&\dots&0\\
  \end{pmatrix} 
\end{align}
\end{widetext}

 such that 
 \begin{equation}
   \label{eq:Hnk}
   H(\{n_k\}) = \hat H_{\rm TLS}+\hat H_B+ \hat V = A+B   
 \end{equation}

 We emphasize that this is the representation of the Hamiltonian in the subspace defined by Eq.\ \eqref{eq:nksubbasis}, which depends on the initial set $\{ n_k \}$. We now investigate the partition function $\Xi_{\{n_k\}} = \tr\{\exp(-\beta H(\{n_k\}))\}$ by approximating the energy eigenvalues in $H(\{n_k\})$ using Weyl's matrix inequalities\cite{bhatia2013matrix}, which we state next in our context. In brief, the eigenvalues of $A$ and $B$ provide lower and upper bounds for the energy eigenvalues in $H(\{n_k\})$ that depend on the inverse density of bath modes $\Delta \omega$.

 Since $A$ and $B$ are $(N+1)$-dimensional Hermitian matrices their eigenvalues, which we will denote by  $\{\alpha_k\}$ and $\{\gamma_k\}$ respectively, can be listed in decreasing order.  Thus we write
 \begin{align}
   \alpha_k =& -\frac{\omega_L}{2}+\omega_{N-k} +\varepsilon_B, \label{eq:alphas}
\end{align}
with  $0 \leq k \leq N$ and $\omega_0 = 0$,  as well as
\begin{align}
   \gamma_0 = & \frac{1}{2}\left(\omega_L+\sqrt{\omega_L^2+4 \eta} \right)\\
   \gamma_N = & \frac{1}{2}\left(\omega_L- \sqrt{\omega_L^2+4 \eta} \right)\\
   \gamma_k =&0  \hspace{0.5cm} \text{otherwise}, \label{eq:gammas}
 \end{align}
where we have introduced the parameter $\eta =(1/4N)\sum_k|u_k|^2$.  In order to obtain $\gamma_i$, we have noticed that the characteristic polynomial $p(\gamma)=\det(B- \gamma I)$ can be evaluated by using the Laplace Expansion Theorem \cite{marcus1965introduction}, and it equals to
\begin{align}
 p(\gamma) =& \sum_{l=1}^N \left[(\omega_L - \gamma)(- \gamma) - \frac{|u_l|^2}{4} \right] (-\gamma)^{N-1}\\
          =& N \left[(\omega_L - \gamma)\gamma + \eta \right] (-\gamma)^{N-1}.
\end{align}

If we denote by ${\lambda_k}$ the eigenvalues for the $H(\{n_k\})$ in Eq.\ \eqref{eq:Hnk} and they are listed in decreasing order, the eigenvalues $\{\alpha_k\}$, $\{\gamma_k\}$ and $\{ \lambda_k \}$ satisfy the following inequalities\cite{bhatia2013matrix}
\begin{align}
\lambda_k \leq&\, \alpha_j + \gamma_{k-j} \hspace{0.5 cm} (j \leq k ) \label{eq:Weylleq}\\
\lambda_k \geq&\, \alpha_j + \gamma_{k-j+N} \hspace{0.5 cm} (j \geq k ), \label{eq:Weylgeq}
\end{align}
 in particular, if $k = j$ then
 \begin{align}
\lambda_k \leq& \alpha_k + \gamma_{0}  \label{eq:ineqmax}\\
\lambda_k \geq& \alpha_k + \gamma_{N} .\label{eq:ineqmin}
 \end{align}
Moreover, if $j = k -1$ from Eq.\ \eqref{eq:Weylleq} we obtain
\begin{align}
  \lambda_k \leq \alpha_{k-1} + \gamma_{1} \label{eq:ineqless}
\end{align}
 and if $j= k +1$ from Eq. \eqref{eq:Weylgeq} we have
 \begin{align}
\lambda_k \geq \alpha_{k+1} + \gamma_{N-1} \label{eq:ineqgreat}.
 \end{align}
From the inequalities in Eqs.\ \eqref{eq:ineqless} and \eqref{eq:ineqgreat} together with the eigenvalues in Eqs.\ \eqref{eq:alphas} and \eqref{eq:gammas} we obtain upper and lower bounds for $\lambda_k$  with $1\leq k \leq N-1$
\begin{align}
  -\frac{\omega_L}{2}+\omega_{N-k-1} +\varepsilon_B \leq \lambda_k \leq -\frac{\omega_L}{2}+\omega_{N-k+1} +\varepsilon_B, \label{eq:ineqlambdak}
\end{align}
and since $\omega_k = k \Delta \omega$, Eq.\ \eqref{eq:ineqlambdak} is equivalent to $|\lambda_k - \alpha_k| \leq \Delta \omega$. Consequently, for small $\Delta \omega$, we can approximate
\begin{equation}
  \label{eq:lambda_approx}
  \lambda_k = \alpha_k,
\end{equation}
 for $1\leq k \leq N-1$. It remains to determine appropriate approximations for $\lambda_0$ and $\lambda_N$. For the former, considering Eq.\ \eqref{eq:ineqmax} 
\begin{align}
  \lambda_1 \leq& \lambda_0 \leq \alpha_0 +\gamma_{0} \\
-\frac{\omega_L}{2}+\omega_{N-1} +\varepsilon_B \leq& \lambda_0 \leq -\frac{\omega_L}{2}+\omega_{N} +\varepsilon_B + \gamma_{0}. 
\end{align}
as the ordering in $\{ \lambda_k\}$ dictates that $\lambda_1 \leq \lambda_0$. Since $\gamma_0$ can take large values in the strong coupling regime, in this case our estimate will be 
\begin{align}
  \lambda_0 =& -\frac{\omega_L}{2}+\omega_{N-1} +\varepsilon_B + C(\Delta \omega + \gamma_0)\\
  =& \alpha_1 + C(\Delta \omega + \gamma_0)
\end{align}
where $0 \leq C \leq 1$ is a constant determined below.   For the latter, in view of Eq.\ \eqref{eq:ineqmin} we find
\begin{align}
 \alpha_N +\gamma_{N} \leq& \lambda_N \leq \lambda_{N-1}  \\
-\frac{\omega_L}{N} +\varepsilon_B +\gamma_{N} \leq& \lambda_N \leq -\frac{\omega_L}{2}+\omega_{1} +\varepsilon_B. 
\end{align}
which suggests that 
\begin{align}
  \lambda_N =& \alpha_{N-1} + C'( \gamma_N - \Delta \omega)
\end{align}
with $0 \leq C' \leq 1$. Finally, we determine the constants $C$ and $C'$ by computing the trace for $H(\{n_k\})$ in Eq.\ \eqref{eq:Hnk}. Indeed,

\begin{align}
  \tr \{ H(\{n_k\}) \} =& \tr\{A \} +\tr \{ B\}\\
 =& (1-N) \frac{\omega_L}{2} +(N+1)\varepsilon_B \notag\\
&+ \Delta \omega \frac{N(N+1)}{2}
\end{align}
On the other hand 
\begin{align}
  \sum_{k=0}^N \lambda_k =&   \alpha_1 +\alpha_{N-1}+ \sum_{k=1}^{N-1} \alpha_k \notag\\
& + C'( \gamma_N - \Delta \omega) +C(\Delta \omega + \gamma_0)\\
=& -(1+N)\frac{\omega_L}{2}+ (N+1)\varepsilon_B \notag \\
& +  \Delta \omega \frac{N(N+1)}{2}\\
&+ C'( \gamma_N - \Delta \omega) +C(\Delta \omega + \gamma_0)
\end{align}
and if $C' = C = 1$, $\sum_{k=0}^N \lambda_k  = \tr \{ H ({n_k}) \}$.

Next, we calculate the partition function for $\hat H _{\{n_k\}}$, $\Xi_{\{n_k\}} = \tr\{\exp(- \beta \hat H _{\{n_k\}} )\}$ is
\begin{align}
   \Xi_{\{n_k\}} =& e^{-\beta \lambda_0}+ e^{-\beta \lambda_N} + \sum_{k=1}^{N-1} e^{-\beta \lambda_k} \label{eq:Xink}\\
 =&  e^{-\beta \lambda_0}-e^{-\beta \alpha_0}+ e^{-\beta \lambda_N} - e^{-\beta \alpha_N} + \sum_{k=0}^{N} e^{-\beta \alpha_k}\\
=& e^{-\beta(\varepsilon_B-\omega_L/2)} \mathcal{R}(\omega_{\rm max}, N)
\end{align}
 where we have introduced the function
 \begin{align}
   \mathcal{R}(\omega_{\rm max}, N) =& e^{-\beta \omega_{\rm max}}\left(e^{-\beta \gamma_0}-1\right) \notag \\
&+ e^{-\beta \gamma_N} - 1 + \sum_{k=0}^{N} e^{-\beta k \Delta \omega}
 \end{align}

In order to recover the canonical partition function we now sum over all families $\{ n_k \}$. Letting $\mathcal{S}$ be such collection we write ($\mu = 0$) 
\begin{align}
  \Xi =& \sum_{\{n_k\} \in \mathcal{S}} \Xi_{\{n_k\}}\\
      =& \left(\sum_{\{n_k\} \in \mathcal{S}} e^{-\beta(\varepsilon_B-\omega_L/2)} \right) \mathcal{R}(\omega_{\rm max}, N) 
\end{align}

We notice that
\begin{align}
  \sum_{\{n_k\} \in \mathcal{S}} e^{-\beta(\varepsilon_B-\omega_L/2)} =& e^{\beta \omega_L/2} \sum_{\{n_k\} \in \mathcal{S}} \prod_{k=1}^N e^{-\beta n_k \omega_k}\\
=& e^{\beta \omega_L/2} \prod_{k=1}^N  \sum_{n=0}^\infty e^{-\beta n \omega_k}\\
=&  e^{\beta \omega_L/2} \prod_{k=1}^N  \frac{1}{1 - e^{-\beta \omega_k}},
\end{align}
and therefore
\begin{align}
  \ln \Xi =& \sum_{k=1}^N \ln \left[\frac{ e^{\beta \omega_L/2}}{1 - e^{-\beta \omega_k}} \mathcal{R}(\omega_{\rm max}, N) \right],
\end{align}
which in the thermodynamic limit leads to the integral form
\begin{align}
 \ln \Xi = \int^{\omega_{\rm max}}\frac{d \omega}{2 \pi} A(\omega) \ln \left[\frac{ e^{\beta \omega_L/2} e^{-\beta \gamma_N}}{1 - e^{-\beta \omega}} \right].
\end{align}
Consequently, the final form for the canonical potential is
\begin{align}
  F=& \frac{1}{\beta} \int \frac{d \omega}{2 \pi} \mathcal{A}(\omega) \ln \left[(1 - e^{-\beta \omega})e^{-\beta \Delta / 2 }\right] 
\end{align}
with $\Delta = \sqrt{\omega_L+4 \eta}$, and that can be further simplified to the form in Eq.\ \eqref{eq:GCP_TLS}.

\section{Spectral density for the damped two-level system}\label{ap:Gr_TLS}

Consider the Green's function
\begin{align}
\mathcal{G}(\tau_2, \tau_1) =&  -i \langle \mathcal{T}_c \hat \sigma^- (\tau_2) \hat \sigma^+ (\tau_1) \rangle   \label{eq:ap_GF_TLS}  .   
\end{align}
The equation of motion for $\hat \sigma^-$ is
\begin{align}
  i \frac{d}{d\tau_2}\hat \sigma^-(\tau_2) =& \omega_L - 2 \sum_k V_k \hat S_z \hat a_k,  
\end{align}
where  $V_k = i u_k / 2$. Then, the equation of motion for the Green Function in Eq.\ \eqref{eq:ap_GF_TLS} is 
\begin{align}
  i \frac{d}{d\tau_2} \mathcal{G}(\tau_2, \tau_1) =& - 2 \delta (\tau_2, \tau_1)
	\langle \hat S_z (\tau_1)\rangle +\omega_L \mathcal{G}(\tau_2\tau_1) 
	\notag\\  &
 - 2 \sum_k V_k\left[ -i \big< \hat S_z(\tau_2) \hat a_k(\tau_2) \hat \sigma^+(\tau_1) \big> \right]  \label{eq:EOM_G_tls} .   
\end{align} 
In order to solve the EOM in Eq.\ \eqref{eq:EOM_G_tls} we approximate the higher order correlation function by the product
\begin{align}
  -i \langle \hat S_z(\tau_2) \hat a_k(\tau_2) \hat \sigma^+(\tau_1)\rangle =& \langle \hat S_z(\tau_2) \rangle \left[ - i \langle \hat a_k(\tau_2) \hat \sigma^+(\tau_1)\rangle \right] \label{eq:ap_approx_Gr}.   
\end{align}
Such decoupling schemes were used in other contexts in Refs.\ \citenum{levy2013steady} and \citenum{bulka2004electronic}. Following the same rationale as in Appendix \ref{ap:GrHO}  we find:
\begin{align}
  -i \langle \hat a_k(\tau_2) \hat \sigma^+(\tau_1)\rangle =& V_k^*\int d\tau' u_k(\tau_2, \tau') \mathcal{G}(\tau', \tau_1)  
\end{align}
which upon substitution in Eq.\ \eqref{eq:EOM_G_tls} leads to the expression
\begin{align}
   i \frac{d}{d\tau_2} \mathcal{G}(\tau_2, \tau_1) =& \delta (\tau_2, \tau_1) \mathcal{S} (\tau_1) +\omega_L \mathcal{G}(\tau_2\tau_1) 
	\notag\\  &
 + \mathcal{S}(\tau_2) \sum_k |V_k|^2 \int d\tau' u_k(\tau_2, \tau') \mathcal{G}(\tau', \tau_1)\label{eq:EOM_G_tls_2}  .  
\end{align}
This equation may be converted to the standard form of the Dyson equation, by introducing the transformation $\mathcal{\tilde G}(\tau_2, \tau_1) = \mathcal{S}^{-1/2}(\tau_2) \mathcal{G}(\tau_2, \tau_1) \mathcal{S}^{-1/2}(\tau_1)$ as shown in Ref. \citenum{ochoa2014non}. As a result we find that in stationary state 
\begin{align}
  \mathcal{G}^r(\omega) = \frac{\mathcal{S}}{(\omega - \omega_L)+ i \Gamma \mathcal{S}/2}.  
\end{align}
From this result we obtain Eq.\ \eqref{eq:DOS_tls}.
  
\section{Nonequilibrium distribution  given by Eq.  (\ref{eq:dist_tls})}\label{ap:noneq_dis_TLS} 

Starting from the gradient expansion employed  in Eq.\ \eqref{eq:grad_ex_ap}, which is valid for $\mathcal{\tilde G}^<$ introduced in the Appendix \ref{ap:Gr_TLS}, and noticing that
 \begin{align}
  \frac{\partial \mathcal{\tilde G}^r}{\partial t} = \dot\omega_L (\mathcal{\tilde G}^r)^2, 
	&&
	\frac{\partial \mathcal{\tilde G}^a}{\partial t} = \dot \omega_L (\mathcal{\tilde G}^a)^2,
\end{align}  
we get
\begin{align}
  \mathcal{\tilde G}^<(t, \omega) =&   \mathcal{\tilde G}^r(t, \omega) \tilde \Sigma^<(\omega)  \mathcal{\tilde G}^a(t, \omega) 
	\notag\\  &
 + i \frac{\dot \omega_L}{2}  \mathcal{\tilde G}^r(t, \omega)  \mathcal{\tilde G}^a(t, \omega)
	\notag\\  &  \times 
 \left\{  \mathcal{\tilde G}^a(t, \omega) -  \mathcal{\tilde G}^r(t, \omega)\right\} 	
\frac{\partial}{\partial \omega} \tilde \Sigma^<(\omega),    
\end{align}
with $\mathcal{A}$ given by Eq.\ \eqref{eq:DOS_tls}. From this result, we recover $\mathcal{ G}^<(t, \omega) = \mathcal{S}^{1/2} \mathcal{\tilde G}^<(t, \omega) \mathcal{S}^{1/2}$, and after some algebraic manipulations we arrive at the expression:
\begin{align}
  i \mathcal{G}^< (t, \omega) = \mathcal{A}(t, \omega)
	\left [n(\omega)+ \frac{\dot \omega_L}{2} \mathcal{S}^{-1} \mathcal{A}(t, \omega) \frac{\partial n(\omega)}{\partial \omega} \right],
\end{align}
which brings the result for $\phi_3(t, \omega)$  given by Eq.\ \eqref{eq:dist_tls}.



\bibliography{BosonThermo}
\end{document}